%% file: main.tex
\documentclass[letterpaper]{article} %
\usepackage[preprint]{aaai2027}
\usepackage[hyphens]{url} %
\usepackage{graphicx} %
\usepackage{natbib} %
\usepackage{caption} %
\usepackage{booktabs}
\usepackage{epsfig}
\usepackage{multirow}
\usepackage{amsmath}
\usepackage{amsfonts}

\usepackage{tikz}
\usetikzlibrary{arrows.meta,positioning,fit,backgrounds,calc}
\input{figures/cast_overview_defs}
\newenvironment{promptlist}{%
  \begin{list}{--}{%
    \setlength{\topsep}{1pt}\setlength{\partopsep}{0pt}%
    \setlength{\itemsep}{0pt}\setlength{\parsep}{0pt}%
    \setlength{\leftmargin}{1.1em}\setlength{\labelwidth}{0.6em}%
    \setlength{\labelsep}{0.5em}\setlength{\rightmargin}{0pt}}%
}{\end{list}}
\newenvironment{promptblock}{%
  \par\nobreak\vspace{5pt}%
  \begingroup
  \scriptsize\raggedright
  \setlength{\parindent}{0pt}\setlength{\parskip}{1pt}%
  \noindent\hrule height 0.4pt\vspace{5pt}%
}{%
  \par\vspace{5pt}\hrule height 0.4pt\endgroup\par\medskip
}
\title{CAST: A Deployed Speech-to-Text Cascade for Closed Captioning of Entertainment Television}
\author{
    Aditya Mishra\equalcontrib, Thanish Bolla\equalcontrib, Bidisha Sharma, Balaji Pitchaikannu, Surbhi Sinha, Anup Kotalwar
}
\affiliations{
    Warner Brothers Discovery\\
    \{aditya.mishra, thanish1.bolla, bidisha.sharma, balaji.p, Surbhi.Sinha, anup.kotalwar\}@wbd.com
}

\begin{document}

\maketitle

\begin{abstract}
Despite rapid progress in automatic speech recognition (ASR), caption generation for long-form entertainment video remains challenging for content dense in music and effects, rich in accents, crosstalk, and noise. The transcript quality depends on segmentation and surrounding modules as much as the acoustic model. A single pass voice activity detection (VAD) can miss faint, low-energy voice while forwarding near-silent segments that induce hallucination and repetition in an autoregressive decoder. These competing errors motivate a hierarchical VAD that revisits rejected segments under semantic and acoustic retention checks. We present Captioning ASR Stack for Television (CAST), an integrated speech-to-text pipeline for closed captioning (CC) combining frozen Whisper large-v3 with segmentation, language identification, lyrics routing, and targeted hallucination detection and repair on a single NVIDIA L4 GPU. On 100 in-house assets spanning 60.46 hours, CAST reduces normalized WER from 6.99\% to 5.36\% and subtitle edit rate from 254.2 to 221.9 min/hour against the WhisperX pipeline, and improves deletion rate by 11\% on public datasets. We also present the behavior of each stage in the pipeline. CAST entered the production speech-to-text (STT) stage of our CC pipeline in late August 2026; we report the evaluation evidence and path to full-scale operation.
\end{abstract}

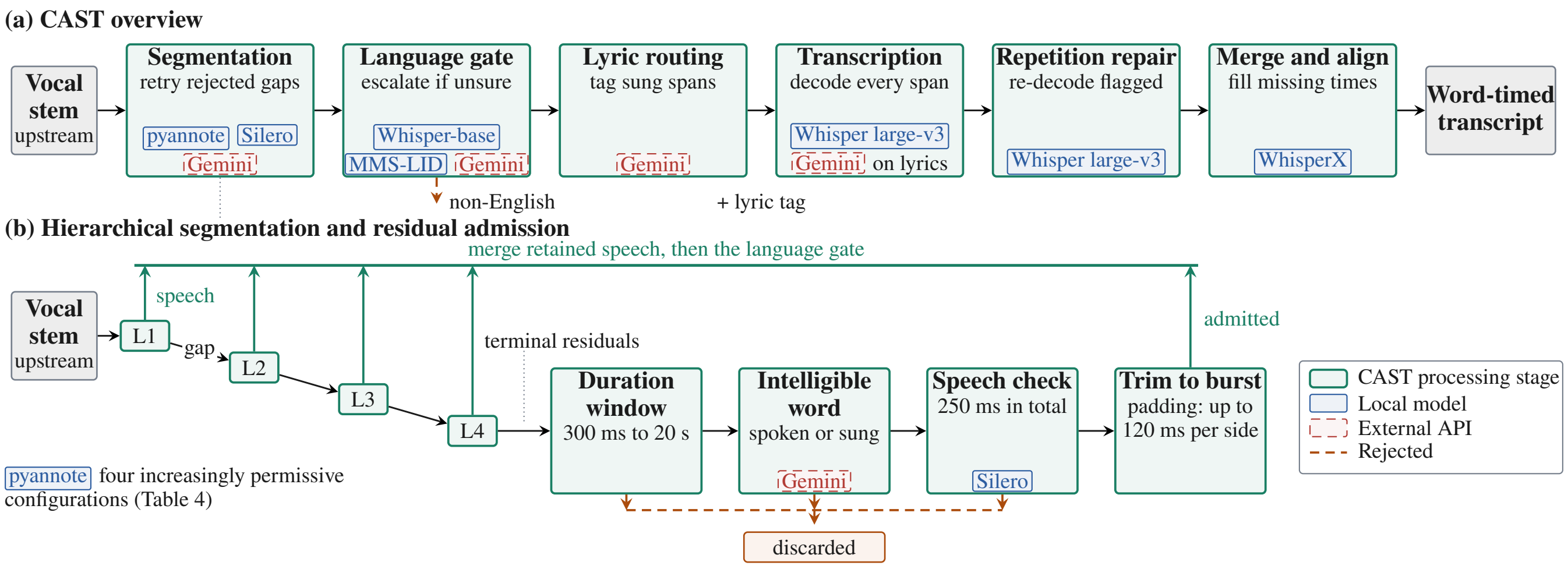
\begin{figure*}[t]
\centering
\input{figures/cast_overview_body}
\caption{(a)The CAST pipeline, (b) the hierarchical segmentation and residual admission stage.}
\label{fig:cast}
\end{figure*}
\vspace{-0.3cm}
\section{Introduction}

Our catalog spans feature films, episodic drama and comedy, unscripted reality,
news, children's animation, live sports, and lifestyle programming, with
correspondingly diverse acoustics. Music, effects, and laughter are deliberately
mixed beneath dialogue, while naturalistic performance adds off-mic speech, rapid
interjections, overlapping turns, and singing that must still be captioned.
Captioning begins from a mixed program master rather than a close-talk
microphone. The ASR component transcribes audio and passes its
output to later stages, where its errors are unrecoverable. A word never emitted
can be assigned no speaker and placed in no caption block, while a phrase
hallucinated over silence propagates into a block a reviewer must locate and
delete. ASR quality bounds the entire CC pipeline.

Recent progress has come from large scale weak supervision and, more recently,
audio-language models. Whisper set the paradigm, training one encoder-decoder on
680,000 hours of weakly labeled multilingual audio for strong zero-shot transfer
without dataset-specific fine-tuning~\cite{radford2023whisper}; newer open-weight
families include Canary~\cite{puvvada2024canary} and Parakeet~\cite{xu2023tdt},
alongside audio-LLMs such as Voxtral~\cite{liu2025voxtral}. On ITV entertainment
series, \citet{cettolo2026fbk} reported Whisper large-v3 better on unaggregated
segments, while Voxtral-Mini-3B benefits from longer aggregated segments and
achieves lowest WER. We adopt Whisper large-v3 for ASR as its permissive open weights let us serve the recognizer inside our own project boundary and control the serving configuration at catalog scale.

As sequence-to-sequence models are trained on short utterances, input
segmentation governs long-form inference~\cite{gaido2021beyond,tsiamas2022shas}.
Whisper's 30-second window forces it to traverse long audio in a sequential
decoding loop governed by thresholds on no speech probability, average
log-probability, and compression ratio~\cite{radford2023whisper}; when these
misfire, the model drifts, loops, or emits fluent text over silence and
music~\cite{koenecke2024careless,wang2025calm}. The subtitling community has independently shown that such errors matter beyond WER, since subtitle quality depends jointly on text, timing, and line segmentation~\cite{wilken2022suber,papi2023direct}. The widely adopted solution, VAD
segmentation with chunking and external forced alignment~\cite{lan2026hwtsc,bain2023whisperx} on full entertainment programmes, addresses both omitted speech and the errors that more permissive detection exposes.

Long-form media ASR gained broad attention with the Multi-Genre Broadcast (MGB) challenges, which extended earlier broadcast-news evaluations to 1{,}600 hours of multi-genre BBC television \cite{bell2015mgb}. WhisperX based ASR pipeline has been established as the most accurate in earlier work~\cite{lucca2025subtitles,bain2023whisperx}. Each of these failures has been studied on its own; segmentation via hybrid and iterative pause search~\cite{gaido2021beyond}, the coupling of VAD errors to ASR insertions and deletions~\cite{apsipa2024cascade}, hallucination suppression applied downstream of a decoder that has already seen non-speech~\cite{whispercd2026}, and speech-in-music detection as a standalone benchmark~\cite{chaudhuri2018ava}, but not as a single system.

In this work, we present CAST, the STT stage of our CC pipeline in production since late August 2026 on unscripted entertainment television. Deployed captioning systems are seldom described in the literature, published works benchmark recognizers~\cite{lucca2025subtitles} or scores subtitle
quality~\cite{jat2024asr}, but rarely opens the component that produces the transcript. To our knowledge, no prior work on a commercial STT component in entertainment captioning reports its routing decisions, its failures, and its measured effect on caption quality.

We make three contributions. \emph{First}, we show how to adapt a frozen, weakly supervised recognizer to this material without fine-tuning, a hierarchical segmenter re-examines apparent non-speech at progressively more permissive settings, terminal residuals face semantic and acoustic checks before admission, a cheap-to-expensive cascade escalates only ambiguous language decisions, lyric-bearing spans take a dedicated transcription route, and a text-side repetition check re-decodes only flagged spans. \emph{Second}, we instrument every stage on a real catalogue, reporting which audio is admitted, how language and lyric decisions route it, and when repair fires; these are behavioural measurements, not annotated prevalence. \emph{Third}, we report paired caption-quality evidence on 100 assets using WER, its component
errors, SubER, and editor time on a subset, together with the alternatives we abandoned and the path from production cutover to full-scale operation.

\vspace{-0.2cm}
\section{Captioning ASR Stack for Television (CAST) }
\label{sec:system}
Figure~\ref{fig:cast}(a) illustrates the CAST pipeline within the CC workflow. The pipeline processes the audio through source separation, segmentation, language
and lyric routing, transcription, repair, and final word alignment. In this section, we describe different stages of CAST and design choices.

\vspace{-0.2cm}

\subsection{Audio Source Separation}
Entertainment assets arrive as mixed program masters in which instrumental music and sound effects are deliberately balanced against dialogue, and every downstream module of our pipeline inherits that interference. We  first separate the mixture with Demucs~\cite{defossez2019demucs}, retaining only the vocal stem, which preserves both spoken dialogue and singing, the content that we must caption, and we discard the music and effects stems. Every later stage of our pipeline operates on this vocal stem. We use separation for recognition alone: we never deliver the separated audio to viewers, so we optimize it for transcription rather than perceptual quality~\cite{syed2025mss}.

\subsection{Hierarchical VAD and Residual Admission}

\label{sec:hierarchical-vad}
Whisper decodes audio in fixed 30 second windows, so long-form assets must be cut
into short segments, and how we cut them determines what the ASR model ever sees.
Missing a low-energy speech region leaves a missing caption block; passing a
silent window to the decoder as speech invites hallucination, which a single-pass VAD
cannot avoid.

In the CAST pipeline, we propose a hierarchical VAD, which has multiple levels with increasing permissiveness. A speech region that the strictest level rejects is not discarded but re-examined at the next level, which typically recovers quiet or heavily masked dialogue; wherever a deeper level finds speech, that region rejoins the ordinary speech path and is treated like any other
utterance. A region that no level accepts becomes a \emph{terminal residual}, which
we forward only when two independent checks agree, a semantic verdict that it
contains transcribable content, and an acoustic energy based filter. Difficult speech therefore gets several chances to reach transcription, while the residual path that produces
hallucinations stays tightly gated. Following we describe the stages of the hierarchical VAD.

\vspace{2pt}\noindent\textbf{Partition:} We consider the long-form audio
as the root node and use Pyannote VAD~\footnote{https://huggingface.co/pyannote/voice-activity-detection} as our base detector. At configuration level \(\ell\) we split a node into the
speech intervals Pyannote detects and the non-speech intervals referred as {\it{residual}} between them, and we keep both as children. Conventional pipeline throws away the residual part.

\noindent\textbf{Recurse:} We treat a speech child as terminal, and we
retry every residual at level \(\ell{+}1\), where we make the schedule
monotonically more permissive (onset \(0.85\!\rightarrow\!0.75\), minimum speech
\(300\!\rightarrow\!50\)~ms, minimum silence \(350\!\rightarrow\!60\)~ms;
Table~\ref{tab:vad-levels}). When a residual yields speech at a finer level, we
return it to the speech path as ordinary detected audio. If a configuration
reproduces the partition of its parent, we consider it to have made no progress
and drop the node to the next level; once we exhaust level~4, we make the node
terminal. We use four levels, which bounds the recursion depth.

\noindent\textbf{Admit:} We consider a terminal residual to be audio
that no VAD level would accept, and we pass it through three gates in series.
The \emph{duration window} keeps only spans between 300~ms and 20~s. The
\emph{intelligible word} gate, our semantic check, then asks Gemini~2.5~Pro for
a schema-constrained binary verdict, which we define as positive only for at
least one intelligible spoken or sung word or a recognizable clipped word-part,
and explicitly negative for humming, breathing, laughter, screams, lyric-less
music, and effects. In the \emph{speech check}, we deploy Silero to detect at least 250~ms of speech in total within the span. For spans that pass, we then apply \emph{trim to burst}, where we keep only the longest detected speech burst, padded by up to 120~ms on each side and bounded by the original span, and we discard separate bursts outside that retained interval. 

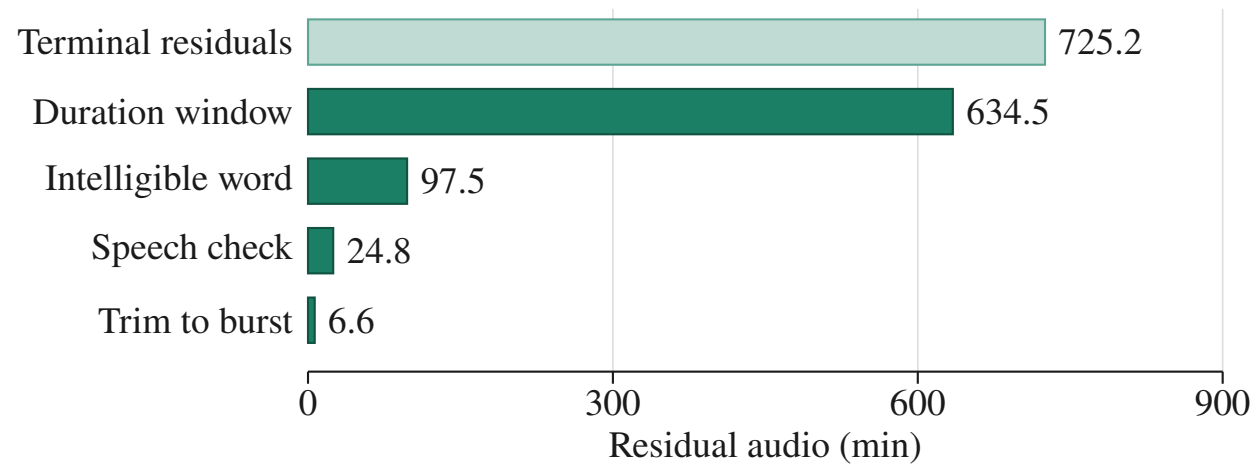
\begin{figure}[!tb]
\centering
\input{figures/vad_admission_body}
\caption{Illustration of residual admission process of the segmentation stage across 100 assets.}
\label{fig:vad-cascade}
\end{figure}

\noindent\textbf{Reassemble:} We merge all the surviving children in successive into less than 30~s segment to keep the neighbouring contextual information. In Figure~\ref{fig:vad-cascade}, we show how we recover the residual audio through admission using different gates.

\subsection{Language Identification}
As we aim to caption English content only through the CAST pipeline, we add a language identification (LID) gate before decoding. As we use English language code in Whisper large-v3 inference, a foreign-language utterance may return plausible English tokens that become substitution errors.

In this stage, we decide which speech regions are English before passing to the transcription stage. Dropping accented English loses captions, while admitting foreign speech adds substitutions. We label each admitted span longer than 650~ms with a cheap-to-expensive cascade, because no single threshold serves both risks and escalation spends compute only on ambiguous spans. Whisper-base is the fast path when the English probability exceeds 0.8; we send the rest to the LID model MMS-LID-4017~\footnote{https://huggingface.co/facebook/mms-lid-4017} and accept its top label only at confidence 0.6 or above. Gemini~2.5~Pro handles low-confidence cases and MMS labels drawn from a curated set of 33 codes that are often returned for accented broadcast English, giving those spans a second decision rather than dropping them on the MMS label alone. We remove spans not labelled as English and we donot provide a separate multilingual transcription path. 

\begin{figure}[t]
\centering
\input{figures/language_filter_body}
\caption{Illustration of language identification stage by assigned language label across 100 assets.}
\label{fig:language-filtered}
\end{figure}
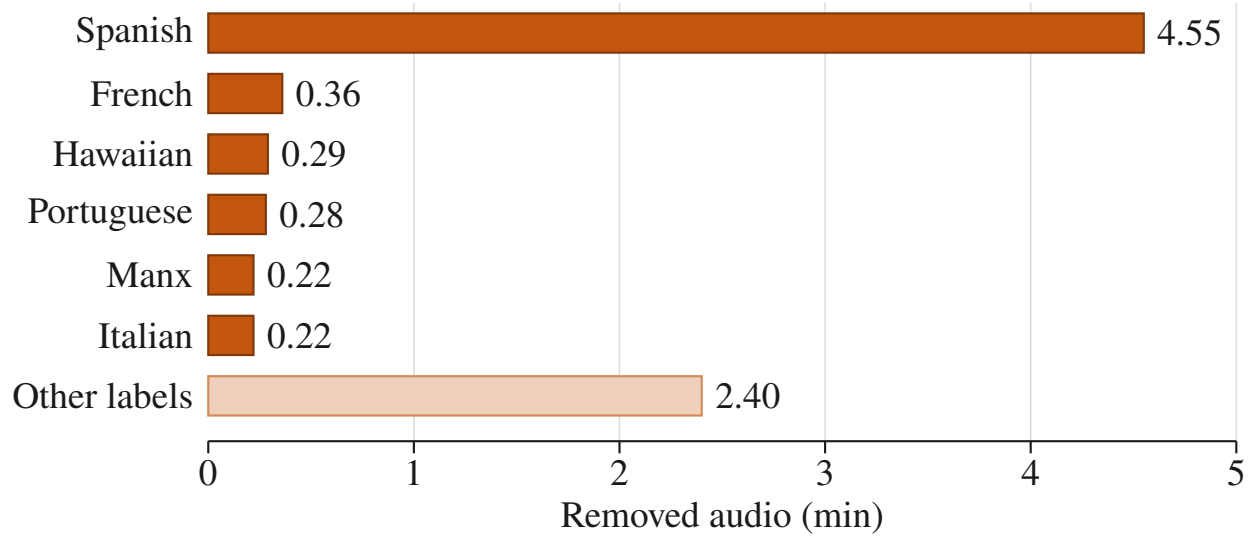

Across the 100 assets, Whisper's fast path resolves 9{,}206 of 9{,}842
segment-level language decisions (93.5\%). MMS resolves another 106, while
530 spans (5.4\%) reach Gemini. Gemini labels 302 of these spans as English
(57.0\%), retaining them at the language gate; the other 228 receive a unknown label and are removed. In Figure~\ref{fig:language-filtered}, we show the six largest named labels by duration, led by Spanish, and groups the remainder across 100 assets.

\subsection{Lyric Classification and Routing}
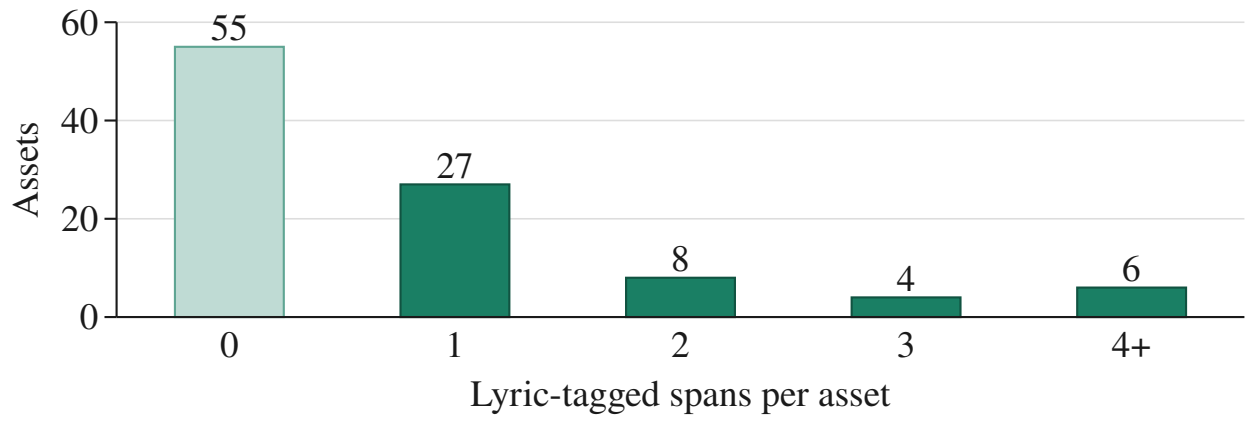
\begin{figure}[!tb]
\centering
\input{figures/lyric_routing_body}
\caption{Statistics of lyrics classification stage of the CAST pipeline across 100 assets.}
\label{fig:music-overview}
\vspace{-0.35cm}
\end{figure}

Media files interleave songs and musical passages with dialogue, and general ASR
models handle them poorly. Sung and rapped lines carry narrative information
and must be captioned. So we add a classification stage that uses Gemini~2.5~Pro to
identify lyric-bearing segments. The classification is narrower than general music detection; singing, rapping, a cappella, jingles, themes, and short lyric-bearing phrases are
positive, while instrumental music, wordless humming, and ordinary speech over a
music bed are negative and a mixed span that is mostly speech stays on the
speech path. We merge consecutive positive spans. We route lyric-tagged segments to the lyrics-oriented transcriber, for which we again use Gemini~2.5~Pro. Figure~\ref{fig:music-overview} shows that most affected assets have only one tagged span (27 of 45), while six have four or more. Lyric aware transcription thus serves a small amount of audio spread across nearly half
the cohort.

\subsection{Transcription}

We decode every speech segment with Whisper large-v3 at temperature zero with the
language fixed to English. Where segments are close in time, we also pass the preceding
transcript as an initial prompt, which carries lexical context across boundaries. We decode lyric-tagged segments with Gemini~2.5~Pro, because the Whisper-large-v3 model under transcribes sung content. We prompt for audible spoken, sung, and rapped words only, including repetitions actually performed so that a chorus survives intact.

In development we observe that Whisper omits words even when the segment contains the complete utterance. Therefore, we impose explicit protection by adding 50~ms of leading silence unless the segment already begins with silence or shows multiple language candidates. After decoding, we run a forced alignment, when the final word occupies an unusually large fraction of the segment (a signature of truncation) we redecode with additional trailing silence. We then pass the transcripts to repetition detection and repair stage.

\vspace{-0.2cm}
\subsection{Repetition Detection and Repair}

If the hierarchical VAD forwards some spurious silence, the decoder may loop or emit characters and enter hallucination stage. Deleting the silence would discard the words the front end recovers, so we re-decode locally before falling back.

In the repetition detection stage, we search for two attributes, a phrase repeated in immediate succession, using \(n\)-grams up to \texttt{max\_ngram}=15, and a single character repeated five or more times. A phrase must recur at least
\texttt{min\_repeats}=6 times before we treat it as a candidate, which is high
enough to spare genuine speech a speaker who says ``no no no'' is not looping. For longer runs we require corroboration from at least one further condition, the phrase recurs eight or more times, the run begins after the first fifth of the transcript, it covers at least 70\% of the words, or a six-repeat run begins after 60\% of the transcript.

Once the detector flags a segment, we repair it from the original audio rather than by editing text, escalating only as far as we need to. We first re-decode the whole window under Whisper's temperature fallback schedule (0.0-1.0), since resampling often escapes a loop that the main pass's greedy decode could not. If the loop survives that decode, we search for a split point. We estimate the loop's onset, quantise it to a one second grid and try that point first. Then
walk back from the segment end in one second steps, decoding each candidate and
accepting the first whose two halves are both clean. If only a trailing second
still loops, we keep the clean prefix; if no split succeeds, we retain the original transcript, so a failed repair never deletes words.

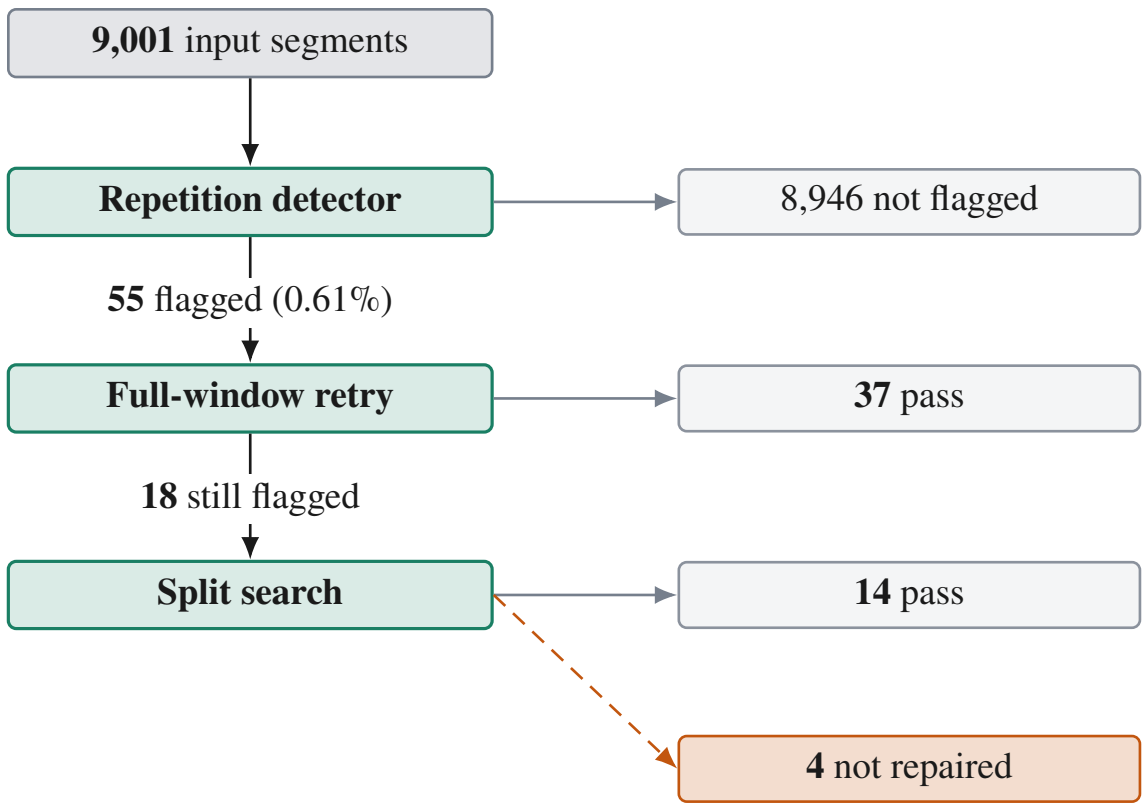
\begin{figure}[t]
\centering
\input{figures/repair_cascade_body}
\caption{Flow of the 55 flagged segments through repetition repair across
100 assets.}
\label{fig:hal-repair}
\vspace{-0.4cm}
\end{figure}

On the 100-asset run, the repetition detector flags 55 of 9{,}001 segments
reaching repair (0.61\%, or 0.91 per audio hour). Figure~\ref{fig:hal-repair}
summarizes the outcomes. A full-window retry produces non-empty output that
passes the repair detector for 37 segments; the remaining 18 enter split search.

\vspace{-0.4cm}

\vspace{0.1cm}
\section{Experiments}
\label{sec:data-evaluation}
In this section, we describe our experimental setup, then the datasets on which we validate CAST, and finally the evaluation metrics through which we assess caption quality along complementary dimensions.
\vspace{-0.2cm}
\subsection{Experimental Setup}
To establish the impact of the multi-stage CAST cascade, we compare it against the
widely used WhisperX pipeline~\cite{bain2023whisperx} on both in-house and
publicly available data. We run GCP infrastructure in development and in
production. On a single NVIDIA L4 GPU, the full cascade occupies roughly
9 to 12~GB of VRAM and averages 27.4 minutes of compute per audio hour. Host RAM
is the binding constraint: programmes up to about 1.5 hours fit a 16~GB host,
while longer assets require 32~GB. Our main infrastructure risk is L4
availability under queued \texttt{FLEX\_START} scheduling.

\vspace{-.2cm}
\subsection{Dataset}~\label{sec:data}
We use two datasets. The first is an in-house corpus that we cannot release. To
support reproducibility, we additionally evaluate CAST on publicly available
data.
\subsubsection{In-house data}
We evaluate WhisperX and CAST on 100 in-house English programme assets
comprising 60.46 hours, 569,651 normalized reference words, and 80,653 reference
caption cues. The cohort spans 48 known series across nine networks, covering
home renovation and property programmes, cooking competitions, factual
entertainment, automotive content, and paranormal series. Network and series
metadata are missing for 22 assets, which we nonetheless retain. Durations range
from 17.6 to 84.6 minutes, with a median of 42.3 minutes. Each asset has a manually
annotated WebVTT file, which we treat as its reference. Annotators mark blocks
containing lyrics or sung audio with musical notes; such blocks appear in 39
assets, across 457 cues.

\subsubsection{Publicly available data}
To make our results reproducible outside our own catalogue, we evaluate CAST on
three public English long-form benchmarks. CORAAL~\cite{kendall2021coraal} provides full-length sociolinguistic
interviews with speakers of African American Language, testing vernacular and
spontaneous speech. Earnings-22~\cite{delrio2022earnings22} is a 125-file,
119-hour corpus of earnings calls from companies in 27 countries grouped into
seven accent regions, and exercises accented multi-speaker audio.
Rev16~\cite{radford2023whisper} comprises the 16 podcast episodes retained from
Rev.ai's 30-episode transcription benchmark, covering conversational studio
recordings with crosstalk. All three are long-form and spontaneous, but none
carries the music and effects beds of a programme master, so they test long-form
robustness rather than our hardest condition. Their references are verbatim
transcripts without cue or line structure, so we report WER alone on these
corpora.

\subsection{Evaluation Metrics}
We use two complementary objective metrics to assess lexical accuracy and the quality
of the resulting timed captions. We also report edit rate as a subjective metric.

\subsubsection{Word error rate (WER)}
For an alignment between hypothesis and reference word sequences,
\begin{equation}
\mathrm{WER}=100\,\frac{S+D+I}{N},
\label{eq:wer}
\end{equation}
where $S$, $D$, and $I$ count word substitutions, deletions, and insertions,
respectively, and $N$ is the number of reference words. We lower-case and
clean punctuation, concatenate each asset's cues, and discard timestamps
before scoring.

\subsubsection{Subtitle edit rate (SubER)}
SubER evaluates subtitle text, line/cue segmentation, and timing jointly
\cite{wilken2022suber}:
\begin{equation}
\mathrm{SubER}=100\,\frac{E_{\mathrm{word}}+E_{\mathrm{break}}+H}
{N_{\mathrm{word}}+N_{\mathrm{break}}},
\label{eq:suber}
\end{equation}
where $E_{\mathrm{word}}$ and $E_{\mathrm{break}}$ count substitutions,
deletions, and insertions of words and break tokens, respectively; $H$ counts
shifts of contiguous token sequences. The denominator counts reference words
and line/cue-break tokens under SubER's normalization. Timing constrains the
alignment, word matches and substitutions require overlapping subtitle cues,
and shifts are also time constrained. We use the default case-insensitive,
punctuation normalized implementation. SubER suits broadcast captioning
because accurate words must also be grouped into readable lines and placed
at appropriate times; captions with identical text can differ in these
properties, which our WER calculation ignores.

\subsubsection{Edit rate}
We additionally measure \emph{edit rate}, a human-effort metric obtained from a post-editing study with professional captioning vendors. Fifteen captioners edited CAST and WhisperX transcripts under blind review, we did not disclose which system produced a transcript, and we assigned the two versions of each asset to different captioners, so no editor saw both versions of the same content. We recorded the editing time of every session and report it as minutes of editing per hour of media.

\vspace{-0.3cm}
\section{Results}
As described in the Dataset section, we evaluate CAST on in-house and public data. Our in-house references are caption files with cue and line structure, so we report WER, SubER, and edit rate. Whereas, the public references are verbatim transcripts, so we report WER alone.
\label{sec:results}

\subsection{Results on In-house Data}
\begin{table}[t]
\centering
\small
\begin{tabular}{lrrr}
\toprule
Metric & WhisperX & CAST & Reduction \\
\midrule
WER (\%) & 6.99 & \textbf{5.36} & 23.31\% \\
SubER (\%) & 14.72 & \textbf{11.18} & 23.31\% \\
Deletions / 1k & 28.79 & \textbf{19.06} & 33.79\% \\
Substitutions / 1k & 22.08 & \textbf{17.20} & 22.12\% \\
Insertions / 1k & 19.07 & \textbf{17.39} & 8.85\% \\
\bottomrule
\end{tabular}
\caption{Evaluation of WhisperX and CAST on in-house data.}
\label{tab:paired-100}
\vspace{-0.3cm}
\end{table}

We first evaluate the WER and SubER metrics on the in-house data for WhisperX and CAST pipelines, which are presented in Table~\ref{tab:paired-100}. We can observe from Table~\ref{tab:paired-100} that CAST reduces WER from 6.99\% to 5.36\% (23.31\% relative reduction) and SubER from 14.72\% to 11.18\% (24.06\% relative
reduction), when compared to WhisperX. It achieves lower WER on 95 assets and lower SubER on 97.

We further analysed insertion, deletion and substitution errors across CAST and baseline WhisperX. The deletions decrease from 16,402 to 10,859, resulting in a 33.79\% relative reduction (28.79 to 19.06 per 1,000 reference words) as observed in Table~\ref{tab:paired-100}. Substitutions and insertions also decrease; therefore, the improved completeness is attributed to lower total lexical error. This pattern is consistent with the purpose of hierarchical VAD and residual admission, where we reconsider audio rejected as non-speech and give brief or faint utterances another opportunity to reach transcription. The counts establish fewer missing reference words at the system level; they do not establish how many contiguous passages were recovered or which stage recovered them. Isolating the VAD contribution requires an otherwise matched on/off ablation. The references are human production feedback captions rather than independently elicited verbatim transcripts. Their editing provenance and possible dependence on production hypotheses limit interpretation of the system difference.

We run the editing study on 35 of the 100 in-house assets (19.9 programme hours). Table~\ref{tab:edit-time} reports edit rate and total edit time. CAST requires 221.9 editing minutes per programme hour against 254.2 for WhisperX, a reduction of 32.3 minutes per programme hour (12.7\%), or 73.7 editor-hours against 84.4 over the group. Per-asset medians are 225.7 minutes per programme hour for CAST and 255.1 for WhisperX, with interquartile ranges of 206.3-235.9 and 236.6-276.3. CAST is faster on all 35 assets with no ties, and a Wilcoxon signed-rank test on the paired differences returns \(p=5.8\times10^{-11}\). Editors therefore spend
roughly 3.7 hours per programme hour on CAST against 4.2 on WhisperX, close to an extra half hour of work for every hour of content under the incumbent transcription.

\begin{table}[t]
\centering
\small
\begin{tabular}{lrrr}
\toprule
Metric & WhisperX & CAST & $\Delta$ \\
\midrule
Edit rate & 254.2 & \textbf{221.9} & --32.3 min/h \\
Total edit time & 84.4 & \textbf{73.7} & --10.7 h \\
\bottomrule
\end{tabular}
\caption{Edit rate and total edit time for 35 assets sampled from the in-house data comparing WhisperX and CAST.}
\label{tab:edit-time}
\vspace{-0.3cm}
\end{table}

\vspace{-0.2cm}
\subsection{Results on Public Data}
We evaluate CAST on CORAAL, Earnings-22 and Rev16, all of which comprise full-length recordings on which the hierarchical segmentation matters. In Table~\ref{tab:public-longform}, we report deletion, substitution and insertion errors and WER for both systems.

From Table~\ref{tab:public-longform}, we can observe that deletions dominate the error budget on these corpora, running at 68.0 to 152.7 per 1,000 reference words for WhisperX, and CAST removes 11.1 to 11.8 of them. 
The trade off differs from the result of in-house data, and demonstrate the significance of the CAST pipeline. More permissive detection buys deletions at the price of insertions, which rise by 3.3 to 6.8 per 1,000 words here but falls by 1.7 on the in-house data. The deletion reduction still exceeds the insertion increase on every corpus. However, in Table~\ref{tab:public-longform} the net WER gain is a modest 0.17 to 0.58 points against 1.63 on the in-house data. We note that deletions matter more than insertions for post-editing time in captioning, as editors must listen to the audio to recover missing words but can remove spurious ones on sight.

\begin{table}[t]
\centering
\scriptsize
\setlength{\tabcolsep}{3.5pt}
\resizebox{\linewidth}{!}{%
\begin{tabular}{lrrrrrrrr}
\toprule
Corpus & \multicolumn{2}{c}{Del/1k} & \multicolumn{2}{c}{Sub/1k}
& \multicolumn{2}{c}{Ins/1k} & \multicolumn{2}{c}{WER (\%)} \\
\cmidrule(lr){2-3}\cmidrule(lr){4-5}\cmidrule(lr){6-7}\cmidrule(lr){8-9}
& WX & CAST & WX & CAST & WX & CAST & WX & CAST \\
\midrule
CORAAL & 152.7 & \textbf{141.7} & 52.4 & 58.0 & 10.0 & 13.3 & 21.52 & \textbf{21.30} \\
Earnings-22 & 68.0 & \textbf{56.2} & 50.7 & 50.6 & 21.7 & 27.8 & 14.04 & \textbf{13.46} \\
Rev16 & 69.8 & \textbf{58.5} & 45.1 & 47.9 & 15.8 & 22.6 & 13.07 & \textbf{12.90} \\
\bottomrule
\end{tabular}}
\caption{Comparison of WhisperX (WX) and CAST on public data.}
\label{tab:public-longform}
\vspace{-0.5cm}
\end{table}

\section{Deployment and Lessons Learned}
\label{sec:deployment}
\subsection{Integration and Serving}
CAST entered production in late August 2026, replacing the previously deployed WhisperX pipeline. In the export step, CAST shapes WhisperX-aligned words into the same keys and nesting consumed from WhisperX, keeping the downstream interface fixed as traffic grows. This is schema-compatible by inspection, not byte-identical output and not a measured claim about integration cost. The STT stage is the first of several. A transcript reviewer edits CAST's output, audio-transcript alignment re-times the reviewed words against the audio, speaker diarization assigns them to speakers, and the result is formatted into closed-caption blocks.

\vspace{-0.2cm}
\subsection{Operational Failure Modes}
In the repetition detection and repair stage of the CAST pipeline, we perform repeated decoding of the same audio. Across the 100 assets in-house data, 18 split searches evaluate 156 candidate splits, with a median of 7 and a maximum of 29. The maximum occurs on a segment of 95~s. Search cost therefore varies substantially even though repair is invoked for only 0.61\% of input segments. We flag this as a limitation of the proposed approach.

\vspace{-0.2cm}
\subsection{Lessons from Development}
The VAD experiment is not the only redesign trigger. Across five repeated runs on two diagnostic assets, pyannote boundaries were stable but Gemini changed its keep/drop judgement on 1.6-2.6\% of residuals. Tightening the prompt and adding Silero corroboration and trimming made merged segment counts identical across the five reruns. Mean WER improved on both assets, but \(n=2\) supports containment, not a general quality claim.

\vspace{-0.2cm}
\subsection{Path to Full-Scale Deployment}
The remaining path to full scale has three gates, staged traffic increases with job-success, latency, and sampled D/I/S monitoring; a documented route back to WhisperX when those gates fail; and mitigation for GPU-queue exhaustion through capacity or regional fallback. Current telemetry contains only about 5 to 10 usable production observations, which is insufficient to estimate editor impact, fleet reliability, or unit economics. We therefore make no
production-outcome claim.

\vspace{-0.2cm}
\section{Limitations}
In this work, we propose CAST, which processes only English content, and compares CAST with one incumbent rather than
external commercial systems. The 100-asset quality results are conditional
on the selection criteria stated under ``Evaluation Data.'' References
are human production-feedback captions; their editing provenance for this
cohort is not independently established. The earlier UAT references were
edited from CAST output and may favour that system. Neither cohort measures
editor time. Stage telemetry comes from a separate instrumented run on the
same 100 source assets, without frame-level speech, language, music, or
hallucination labels. The residual path and repair stage lack on/off ablations, so whole-system error reductions cannot be assigned to either module. Deletion counts also do not measure the number or duration of missing passages. Overlapping-speaker transcription is out of scope, and the recent production cutover leaves operational impact insufficiently measured.
\vspace{-0.2cm}
\section{Conclusion}
In this work, we present the STT component for captioning long-form unscripted television with a frozen recognizer, on the hypothesis that transcript quality also depends on the pipeline around the model. CAST puts that premise into practice, a recall first hierarchical segmenter reconsiders audio that a single VAD pass discards, semantic and acoustic gates decide which of that audio is worth decoding, and the language, lyric, and repetition stages handle the failure modes that more permissive detection exposes.

On 100 in-house assets, CAST reduces WER from 6.99\% to 5.36\% and SubER from
14.72\% to 11.18\% against WhisperX, cutting deletions by
33.79\% with no aggregate rise in substitutions or insertions. The deletion
reduction reproduces on CORAAL, Earnings-22, and Rev16, though with smaller WER
gains and higher insertion counts against verbatim references. In a subjective study, the professional captioners spent 12.71\% less time editing CAST transcripts.

What we cannot yet claim matters as well. Our stage telemetry describes routing
rather than verified recovery, and without matched on/off ablations we cannot
attribute the deletion reduction to any single module. Those ablations, aligned
missing passage review, and editor outcomes measured across a monitored traffic
ramp are the evidence we still owe. For practitioners facing similar material, the transferable part is not our thresholds but the shape of the solution, leave the foundation model alone and spend the engineering effort on segmentation decisions, escalate from cheap to expensive models only on ambiguous input, and make failure asymmetric. So that uncertainty forfeits audio rather than inventing words. In captioning, a missing word costs an editor more than a spurious one, and a pipeline should be built to reflect that.

\bibliography{references}

\appendix
\setcounter{secnumdepth}{1}

\section{Experimental Settings}
\noindent\textbf{Instrumented telemetry run.} All stage analyses use the
completed active checkpoint attempt for each of the same 100 assets selected
for caption-quality evaluation, totalling 60.46 hours. Asset identifiers,
source-video and reference mappings, and completion records match the fixed
selection manifest one-to-one. All 100 attempts share the same recorded
source, behaviour-configuration, model, and prompt fingerprints. Stage
statistics are derived from decision tables, with per-asset identities and row counts
checked against stage summaries. Candidate durations use unique intervals;
residual duration after trimming is computed from retained final boundaries.

The instrumented run has no independent frame-level speech, language, music,
or hallucination labels, so these analyses do not estimate detector precision
or recall. Earlier pilot evaluations remain separate supporting evidence.

\noindent\textbf{Segmenter configuration} Table~\ref{tab:vad-levels} gives the
four pyannote configurations the hierarchy walks, together with the detector
time each level consumed on the 100-asset instrumented run. The levels make
100, 23{,}800, 24{,}651, and 24{,}875 attempts, respectively; 100, 2{,}410, 755,
and 674 are accepted. These are attempts, not unique audio intervals. The total
excludes residual classification and other work outside the detector timers. The
schedule loosens monotonically on all four parameters, so a node retried at a
deeper level is always offered a more permissive detector than the one that
rejected it. Level~3 is the exception to the round numbers: its values are the
output of an earlier threshold-tuning run whose configuration was not
preserved, and we report them as shipped rather than presenting them as a
designed operating point. Whole-node speech detection uses the same pyannote
segmentation model (\texttt{pyannote/segmentation@Interspeech2021}) at every
level; only the post-processing thresholds change.

\begin{table}[h]
\centering
\small
\setlength{\tabcolsep}{4pt}
\begin{tabular}{lrrrrr}
\toprule
& \multicolumn{2}{c}{\textbf{Probability}} & \multicolumn{2}{c}{\textbf{Minimum (s)}} & \\
\cmidrule(lr){2-3}\cmidrule(lr){4-5}
\textbf{Level} & \textbf{onset} & \textbf{offset} & \textbf{speech} & \textbf{silence} & \textbf{Compute} \\
\midrule
1 & 0.850 & 0.550 & 0.300 & 0.350 & 0.48~h \\
2 & 0.820 & 0.500 & 0.160 & 0.200 & 0.84~h \\
3 & 0.810 & 0.481 & 0.055 & 0.098 & 0.84~h \\
4 & 0.750 & 0.400 & 0.050 & 0.060 & 0.83~h \\
\bottomrule
\end{tabular}
\caption{The four pyannote configurations walked by the hierarchical
segmenter, with summed detector time at each level over 100 assets
(60.46 programme hours) on a single Tesla T4 per asset. Rounded level times
need not sum to the rounded total.
Onset and offset are speech-probability thresholds; the minimum columns are the
shortest speech and silence runs the post-processor will keep. Level~3's
non-round values are inherited from an unrecorded tuning run and are reported
as shipped.}
\label{tab:vad-levels}
\end{table}

\noindent\textbf{Residual admission} Residual spans shorter than 300~ms or
longer than 20~s bypass the semantic gate: in the 100-asset run these were
4{,}694 and 163 spans respectively, leaving 22{,}384 of 27{,}241 residuals
eligible. Eligible spans are classified ten at a time with three rate-limit
retries. Silero corroboration runs at threshold 0.5, treats 100~ms of silence
as a burst boundary, discards bursts under 50~ms, requires 250~ms of voiced
audio in total, and pads the retained burst by 120~ms on each side. Every
one of the 22{,}384 residual-classification calls returned successfully, so
the reported funnel reflects model decisions rather than infrastructure failures. Both Gemini gates
run at temperature~0, top-\(p=0.95\), and seed~42 with schema-constrained JSON;
Table~\ref{tab:residual-parameters} lists the shipped values together.

\noindent\textbf{Language-decision accounting} Counts refer to one recorded
language decision per candidate span, before subsequent segment coalescing and
cleanup. Decision intervals are positive and non-overlapping within each
asset. Rejected duration sums the intervals marked for removal, and its
programme share uses the full 60.46-hour source duration. The 530 Gemini
decisions match 530 successful fallback-call records; the remaining 9{,}312
decisions are resolved locally by Whisper or MMS.

\noindent\textbf{Serving configuration} The 100-asset checkpoint run uses
one NVIDIA Tesla T4 and one transcription worker per asset. Earlier
serving studies and pilot evaluations used the single-worker L4 layout.
Hardware-specific timings are not directly comparable across these runs.
Whisper's cross-segment prompt chain resets at worker boundaries, so comparative
runs must hold worker count fixed. The caption-quality comparison uses the
completed paired outputs described under ``Evaluation Data.''

\section{Reproducibility Details}

\subsection{Residual Admission Parameters}

\begin{table}[h]
\centering
\small
\begin{tabular}{lll}
\toprule
\textbf{Gate} & \textbf{Parameter} & \textbf{Value} \\
\midrule
Length & Minimum residual duration & 300~ms \\
Length & Maximum residual duration & 20~s \\
Model & Parallel classification calls & 10 \\
Model & Rate-limit retries & 3 \\
Corroboration & Silero threshold & 0.5 \\
Corroboration & Minimum voiced burst & 50~ms \\
Corroboration & Silence used to split bursts & 100~ms \\
Corroboration & Total voiced audio required & 250~ms \\
Trim & Guard padding on each side & 120~ms \\
\bottomrule
\end{tabular}
\caption{Shipped residual-admission settings.}
\label{tab:residual-parameters}
\end{table}

\subsection{Prompt-Defined Gemini Gates}

Both audio classifiers use Gemini~2.5 Pro with temperature 0, top-\(p=0.95\),
seed 42, and schema-constrained JSON. The prompts below are reproduced with
their shipped wording; only line wrapping and typography have changed.

\noindent\textbf{Residual-admission system prompt}
\begin{promptblock}
You are a strict audio validator for an Automatic Speech Recognition (ASR)
system.

Your job is to decide whether the audio clip contains any intelligible human
speech OR sung lyrics---that is, actual spoken/sung WORDS (or parts of words),
not merely voiced sound.

Context: these clips are produced by an imperfect segmenter, so speech or
lyrics may be incomplete, clipped, or appear only at the very start or end of
the clip.

First, silently identify any words you can actually hear (even a single one).
Then decide:

Return \texttt{contains\_speech} = true if ANY of the following is true:
\begin{promptlist}
\item The clip contains clear, intelligible human speech or sung lyrics (one
real word is enough).
\item A word is cut off or clipped (at the start or end of the clip). A partial
or incomplete word COUNTS as speech, as long as it is recognizably part of a
real spoken or sung word.
\item The speech/lyrics appear only briefly, including at the very start or the
very end of the clip.
\end{promptlist}

The words do NOT need to form a complete or grammatical phrase. A single
intelligible word, or a clearly recognizable part of a word (a clipped word
onset or offset), is enough.

Return \texttt{contains\_speech} = false when the clip contains only
non-lexical audio, i.e. no recognizable word or word-part. Do NOT count any of
the following as speech:
\begin{promptlist}
\item Silence, room tone, or background noise.
\item Non-lexical vocal sounds: humming, breathing, sighs, coughs,
throat-clearing, laughter, screams, or a sustained vowel/``mmm'' sound with no
discernible word.
\item Music, instruments, or singing with no discernible lyrics.
\item Applause, sound effects, or other environmental sounds.
\end{promptlist}

If you cannot identify at least one word or clear part of a word, return false.

Output ONLY valid JSON. If speech or sung lyrics exist:
\texttt{\{"contains\_speech": true\}} Otherwise:
\texttt{\{"contains\_speech": false\}}
\end{promptblock}

\noindent\textbf{Lyric-classification system prompt}
\begin{promptblock}
You are an expert audio classification engine. Your sole task is to analyze
the provided audio file and determine if it contains musical vocals (lyrics)
that are SUNG or RAPPED and can be transcribed.

The single most important signal is MELODIC PITCH: do the words carry musical
pitch (a tune / notes), or a clear rap cadence? If yes -> TRUE. If the words
are simply spoken (flat intonation), even loudly or rhythmically -> FALSE.

DECISION CRITERIA FOR TRUE: set \texttt{has\_music\_lyrics} to TRUE if ANY
portion of the clip contains:
\begin{promptlist}
\item Singing (solo, harmonies, backing vocals, or choir)---words carried on a
melody / musical pitch.
\item Rapping---words delivered with a clear rhythmic rap cadence.
\item A-CAPPELLA / UNACCOMPANIED singing counts as TRUE. Singing with NO
instrumental backing is still TRUE; absence of instruments does NOT make it
FALSE.
\item JINGLES and THEME SONGS: a short sung advertising jingle, a TV/show theme
song, or a sung slogan/tagline counts as TRUE---the words carry a clear melody
even if it is only a few words.
\item A BRIEF sung or rapped passage counts as TRUE, even if it is only a few
words or a single short phrase, and even if the rest of the clip is silence or
instrumental---UNLESS a SPEECH-DOMINANCE OVERRIDE below applies. If you hear
ANY clearly melodic sung phrase or clearly rapped phrase anywhere in the clip,
the answer is TRUE unless overridden.
\end{promptlist}

DECISION CRITERIA FOR FALSE: set \texttt{has\_music\_lyrics} to FALSE if the
audio contains NO sung melody and NO rap cadence anywhere, i.e. it is only:
\begin{promptlist}
\item Purely instrumental music (beats, classical, EDM, with no sung/rapped
vocal line).
\item Spoken narration, conversational speech, dialogue, podcasts, interviews,
shouting, or announcing---EVEN IF music is playing underneath, and EVEN IF the
speech is loud, emphatic, excited, or rhythmic. Talking/shouting over a beat is
NOT music lyrics unless the words actually carry a sung melody or rap cadence.
\item Humming, vocalizations, or voice-like sounds with no discernible
sung/rapped words.
\item Sound effects, ambient noise, or nature sounds.
\item Complete silence.
\end{promptlist}

SPEECH-DOMINANCE OVERRIDES (these take PRECEDENCE over the TRUE criteria
above):
\begin{promptlist}
\item INTRO MELODY BEFORE NARRATION: If a voice briefly hums or sings a simple
short melody at the start (or as a lead-in) and the main content that follows
is spoken narration, classify the segment as FALSE (speech).
\item MAJORITY SPEECH: If more than 50\% of the segment's duration is spoken
narration/conversation, classify the segment as FALSE (speech), even if a
shorter sung or rapped passage is also present.
\end{promptlist}

TIE-BREAKER:
\begin{promptlist}
\item If a voice carries a clear sung MELODY (pitched musical notes) or clear
rap cadence for even a brief moment, classify it TRUE---unless a
SPEECH-DOMINANCE OVERRIDE applies.
\item Choose FALSE only when every voice is flat/spoken with no melodic pitch
and no rap cadence, or when a SPEECH-DOMINANCE OVERRIDE applies.
\end{promptlist}

Output ONLY valid JSON with no preamble, explanation, or markdown formatting.
Format exactly: \texttt{\{"has\_music\_lyrics": true/false, "reasoning": "one
concise sentence describing exactly what you heard that led to this
decision"\}}
\end{promptblock}
\subsection{Transcription Settings}

\noindent\textbf{Decoding and context} Audio is processed at 16~kHz in mono.
Whisper large-v3 uses the English language code and temperature zero. Segments
are processed chronologically within each worker. When the gap from the
preceding segment is at most 200~ms, the last 80 whitespace-separated words of
the most recent non-empty transcript are supplied as the initial prompt.
Longer gaps disable prompting. Context is initialized independently for each
worker.

\noindent\textbf{Initial-silence handling} Before decoding, leading silence
is estimated using RMS amplitude over successive 5~ms windows, with a threshold
of 0.01. Whisper-base supplies language candidates whose probabilities exceed
0.05. The pipeline requests 50~ms of leading silence only when no leading
silence is detected and at most one language exceeds this threshold.
Otherwise, no leading silence is requested. The initial pass requests no
trailing silence.

\noindent\textbf{Final-word diagnostic and retries} After Whisper decoding,
language probabilities are recomputed on the decoder input. When at most one
language exceeds 0.05, an English WhisperX aligner estimates word timings. Let
\(T\) denote the input duration and \(s_K,e_K\) the start and end of the last
word with valid timestamps. The diagnostic computes
\[
r=\frac{e_K-s_K}{T}.
\]
A value above 0.30 triggers up to two additional decodes, requesting 50~ms and
then 100~ms of trailing silence. Retries use temperature zero without the
preceding-segment prompt. Each candidate is aligned against the pre-retry audio
and accepted if its ratio falls below 0.30; otherwise, the initial transcript
is retained. If no word has valid timestamps, the diagnostic does not trigger
a retry. These temporary alignments are discarded; final word alignment follows
repetition repair and segment merging.

\noindent\textbf{Lyric transcription} Lyric-tagged segments additionally use
Gemini~2.5~Pro, with temperature 0, top-\(p\) 1, top-\(k\) 1, seed 42, a
thinking budget of 512 tokens, and a maximum output budget of 10,240 tokens.
Audio is supplied as an inline WAV, and the response schema requires a JSON
object containing a transcription string. The prompt requests verbatim audible
speech, singing, and rap; preserves performed repetitions; prohibits completion,
translation, and paraphrasing; and represents unintelligible words with
``[?]''. Clips without transcribable vocals are instructed to return an empty
string. Gemini's text is selected for lyric-tagged segments and subsequently
passes through repetition detection and repair.

\subsection{Repetition Repair Settings}

Hallucination repair uses \texttt{max\_ngram=15}, \texttt{min\_repeats=6}, and
Whisper's temperature fallback sequence from 0.0 through 1.0.

For repair windows longer than 30 seconds, the wrapper calls the same inner
Whisper transcriber once and joins its native segment texts into one flat
string; the separate overlapping-window method present in the class is not
invoked by this path. If the full-window result still loops, the repair search
tests the estimated onset and then every remaining one-second boundary from the
end backward. Thus a window of duration \(L\) has \(O(L)\) candidate splits,
each re-decoding approximately \(L\) seconds across its two halves.

Repair accounting uses the initial decision, retry-attempt, and terminal
resolution records from the same 100 active checkpoints. Each of the 55 flags
has one full-window retry. A full-window pass requires non-empty output and a
negative repair-detector result; the empty retry is reported separately. The
18 searches contribute 156 split-candidate records, while three explicit
fallback records identify retained original text. Initial retries and fallback
records are excluded from split-search counts. Terminal outcomes sum to all
55 flags, with no cases omitted.

\section{Supporting Pilot Evaluations}
\subsection{Aggregate Pilot Findings}
In a user acceptance testing (UAT), text compared CAST and WhisperX on the same client-supplied audio and reference captions. Eight titles contain 277 minutes and 41,579
reference words. The references were edited from CAST output, which may favour CAST; the magnitude of this reference bias is unknown.

\begin{table}[t]
\centering
\small
\begin{tabular}{lrr}
\toprule
\textbf{Metric} & \textbf{WhisperX} & \textbf{CAST} \\
\midrule
WER, punctuation/casing (\%) & 13.46 & \textbf{9.59} \\
WER, LCNP (\%) & 8.73 & \textbf{6.91} \\
Substitutions / 1k, LCNP & 29.2 & \textbf{24.4} \\
Deletions / 1k, LCNP & 36.7 & \textbf{21.5} \\
Insertions / 1k, LCNP & \textbf{21.5} & 23.2 \\
Punctuation edits / 1k & 25.6 & \textbf{14.0} \\
Missing cues / 1k & 3.5 & \textbf{2.2} \\
Timing-drift edits / 1k & \textbf{41.7} & 55.3 \\
\bottomrule
\end{tabular}
\caption{Pooled UAT results on eight titles. Edit counts are algorithmic
proxies per 1,000 reference words, not observed editing time.}
\label{tab:uat-results}
\end{table}

CAST reduced LCNP deletions by 15.2 per 1,000 words while insertions rose by
1.7 and substitutions fell by 4.8. It achieved lower LCNP WER on seven of
eight titles. The exception, \emph{My Lottery Dream Home}, contained heavy
crosstalk and accents and showed higher substitutions (46.3 versus 33.9 per
1,000) and WER (15.85\% versus 12.89\%). Edit tracking is computed by the
evaluation toolkit rather than observed in live editing sessions. It suggests
fewer word, punctuation, missing-cue, split, and split-merge corrections, but
more timing adjustments; we therefore call it an editor-effort proxy.

Listening checks also expose reference limits. In one \emph{House Hunters
International} passage, CAST recovered the softly spoken phrase ``is so
important,'' which the delivered caption omitted; the scorer therefore counted
correctly recovered words as insertions. Such examples qualify the aggregate
insertion count but do not erase CAST's genuine fabrications.

\subsection{Per-Title UAT Results}

\begin{table}[ht]
\centering
\scriptsize
\setlength{\tabcolsep}{2pt}
\begin{tabular}{p{0.45\columnwidth}rrl}
\toprule
\textbf{Title} & \textbf{WhisperX} & \textbf{CAST} & \textbf{Winner} \\
\midrule
Pull Up To The Palm Beaches & 7.57\% & 2.43\% & CAST \\
Mystery at Blind Frog Ranch & 6.45\% & 4.34\% & CAST \\
Ghost Adventures & 5.51\% & 4.71\% & CAST \\
Renovation Resort Showdown & 6.76\% & 5.13\% & CAST \\
Paranormal Caught On Camera & 6.41\% & 5.32\% & CAST \\
House Hunters International & 6.90\% & 5.83\% & CAST \\
Belle Collective (Season 5) & 15.46\% & 11.27\% & CAST \\
My Lottery Dream Home & 12.89\% & 15.85\% & WhisperX \\
\bottomrule
\end{tabular}
\caption{Per-title UAT WER(\%) after lower-casing and removing punctuation.
Two additional titles with reference-coverage artefacts were excluded from
pooled analysis.}
\label{tab:per-title-results}
\end{table}
\FloatBarrier

\clearpage

\end{document}

%% file: figures/cast_overview_defs.tex
\definecolor{cink}{HTML}{1A1A1A}%
\definecolor{cnavy}{HTML}{2C5CA0}%
\definecolor{cgreen}{HTML}{1A7F64}%
\definecolor{corange}{HTML}{C2560F}%
\definecolor{cred}{HTML}{B23A34}%
\definecolor{cgray}{HTML}{767F8C}%
\newcommand{\Ftl}[1]{{\fontsize{8}{8.8}\selectfont\bfseries #1\par}}%
\newcommand{\Fbdy}[1]{{\fontsize{7}{7.8}\selectfont #1\par}}%
\newcommand{\Fedg}[1]{{\fontsize{7}{7.8}\selectfont #1}}%
\newcommand{\Fchips}[1]{{\fontsize{7}{9.4}\selectfont #1\par}}%
\newcommand{\Fpnl}[1]{{\fontsize{8}{8.8}\selectfont\bfseries #1}}%
\tikzset{%
  chipbase/.style={rounded corners=0.7pt, inner xsep=1.2pt, inner ysep=0.9pt,%
                   font=\fontsize{7}{7.8}\selectfont, line width=0.4pt},%
  chiplocal/.style={chipbase, fill=cnavy!8, draw=cnavy, text=cnavy},%
  chipapi/.style={chipbase, fill=cred!5, draw=cred, text=cred, densely dashed},%
}%
\newcommand{\ChipL}[1]{\tikz[baseline=(c.base)]{\node(c)[chiplocal]{#1};}}%
\newcommand{\ChipA}[1]{\tikz[baseline=(c.base)]{\node(c)[chipapi]{#1};}}%
\newcommand{\Cs}{\hspace{2.5pt}}%
\newif\ifFnum\Fnumfalse%
\newcommand{\Fstage}[2]{\parbox[c][1.50cm][s]{2.12cm}{\vspace{1pt}\centering #1\vfill #2\vspace{0.5pt}}}%
\newcommand{\Fgate}[2]{\parbox[c][1.42cm][s]{1.706125cm}{\vspace{1pt}\centering #1\vfill #2\vspace{0.5pt}}}%
\newcommand{\Fdata}[1]{\parbox[c][1.00cm][c]{0.970cm}{\centering #1}}%
\newcommand{\Fdout}[1]{\parbox[c][1.00cm][c]{1.470cm}{\centering #1}}%

%% file: figures/cast_overview_body.tex
\begin{tikzpicture}[x=1cm, y=1cm, line width=0.4pt, >=stealth]%
\tikzset{%
  stg/.style={rounded corners=1.2pt, draw=cgreen, line width=0.7pt,%
              fill=cgreen!7, text=cink, inner sep=0pt},%
  dat/.style={rounded corners=1.2pt, draw=cgray, line width=0.5pt,%
              fill=black!7, text=cink, inner sep=0pt},%
  ours/.style={rounded corners=1.2pt, draw=cgreen, line width=0.7pt,%
               fill=cgreen!7, text=cink, inner sep=1.2pt,%
               font=\fontsize{7}{7.8}\selectfont},%
  flow/.style={->, draw=cink, line width=0.55pt},%
  keep/.style={->, draw=cgreen, line width=0.7pt},%
  drop/.style={->, draw=corange!88!black, line width=0.7pt, densely dashed},%
  hint/.style={draw=cgray, line width=0.4pt, densely dotted},%
  lbl/.style={fill=white, inner sep=0.6pt,%
              font=\fontsize{7}{7.8}\selectfont, text=cink},%
  free/.style={font=\fontsize{7}{7.8}\selectfont, text=cink, inner sep=0pt},%
}%
\node[free, anchor=west] at (0,1.0145) {\Fpnl{(a) CAST overview}};%
\node[dat] (in) at (0.5696,0)
  {\Fdata{\Ftl{Vocal\\stem}\Fbdy{upstream}}};%
\node[stg] (s1) at (2.4535,0)
  {\Fstage{\Ftl{Segmentation}\Fbdy{retry rejected gaps}}%
          {\Fchips{\ChipL{pyannote}\Cs\ChipL{Silero}\\\ChipA{Gemini}}}};%
\node[stg] (s2) at (4.9125,0)
  {\Fstage{\Ftl{Language gate}\Fbdy{escalate if unsure}}%
          {\Fchips{\ChipL{Whisper-base}\\\ChipL{MMS-LID}\Cs\ChipA{Gemini}}}};%
\node[stg] (s3) at (7.3715,0)
  {\Fstage{\Ftl{Lyric routing}\Fbdy{tag sung spans}}%
          {\Fchips{\ChipA{Gemini}}}};%
\node[stg] (s4) at (9.8305,0)
  {\Fstage{\Ftl{Transcription}\Fbdy{decode every span}}%
          {\Fchips{\ChipL{Whisper large-v3}\\\ChipA{Gemini}\Cs\Fedg{on lyrics}}}};%
\node[stg] (s5) at (12.2895,0)
  {\Fstage{\Ftl{Repetition repair}\Fbdy{re-decode flagged}}%
          {\Fchips{\ChipL{Whisper large-v3}}}};%
\node[stg] (s6) at (14.7485,0)
  {\Fstage{\Ftl{Merge and align}\Fbdy{fill missing times}}%
          {\Fchips{\ChipL{WhisperX}}}};%
\node[dat] (out) at (16.8825,0)
  {\Fdout{\Ftl{\mbox{Word-timed}\\transcript}}};%
\draw[flow] (in) -- (s1);%
\draw[flow] (s1) -- (s2);%
\draw[flow] (s2) -- (s3);%
\draw[flow] (s3) -- (s4);%
\draw[flow] (s4) -- (s5);%
\draw[flow] (s5) -- (s6);%
\draw[flow] (s6) -- (out);%
\draw[drop] (s2.south) -- (4.9125,-1.06);%
\node[free, anchor=west] at (5.05,-1.06) {non-English};%
\node[free] at (8.601,-1.06) {+ lyric tag};%
\draw[hint] (s1.south) -- (2.4535,-1.24);%
\node[free, anchor=west] at (0,-1.36)
  {\Fpnl{(b) Hierarchical segmentation and residual admission}};%
\begin{scope}[yshift=-1.76cm]%
\draw[draw=cgreen, line width=0.9pt] (1.48,0) -- (13.56,0);%
\node[free, text=cgreen] at (7.52,0.17)
  {merge retained speech, then the language gate};%
\node[dat] (vs) at (0.5696,-0.80) {\Fdata{\Ftl{Vocal\\stem}\Fbdy{upstream}}};%
\node[ours, minimum width=0.55cm, minimum height=0.34cm] (v1) at (1.60,-0.80) {L1};%
\node[ours, minimum width=0.55cm, minimum height=0.34cm] (v2) at (2.84,-1.16) {L2};%
\node[ours, minimum width=0.55cm, minimum height=0.34cm] (v3) at (4.08,-1.52) {L3};%
\node[ours, minimum width=0.55cm, minimum height=0.34cm] (v4) at (5.32,-1.88) {L4};%
\draw[flow] (vs) -- (v1);%
\draw[flow] (v1) -- (v2);%
\draw[flow] (v2) -- (v3);%
\draw[flow] (v3) -- (v4);%
\node[lbl] at (2.22,-0.98) {gap};%
\draw[keep] (v1.north) -- (v1.north|-0,0);%
\draw[keep] (v2.north) -- (v2.north|-0,0);%
\draw[keep] (v3.north) -- (v3.north|-0,0);%
\draw[keep] (v4.north) -- (v4.north|-0,0);%
\node[free, anchor=west, text=cgreen] at (1.72,-0.36) {speech};%
\node[free, anchor=west] at (5.45,-0.85) {terminal residuals};%
\draw[hint] (5.9045,-0.97) -- (5.9045,-1.78);%
\node[stg] (g1) at (7.06706,-1.88)
  {\Fgate{\Ftl{Duration\\window}\Fbdy{300 ms to 20 s}}{}};%
\node[stg] (g2) at (9.20219,-1.88)
  {\Fgate{\Ftl{Intelligible\\word}\Fbdy{spoken or sung}}{\Fchips{\ChipA{Gemini}}}};%
\node[stg] (g3) at (11.33731,-1.88)
  {\Fgate{\Ftl{Speech check}\Fbdy{250 ms in total}}{\Fchips{\ChipL{Silero}}}};%
\node[stg] (g4) at (13.47244,-1.88)
  {\Fgate{\Ftl{Trim to burst}\Fbdy{padding: up to\\120 ms per side}}{}};%
\draw[flow] (v4) -- (g1);%
\draw[flow] (g1) -- (g2);%
\draw[flow] (g2) -- (g3);%
\draw[flow] (g3) -- (g4);%
\draw[keep] (g4.north) -- (g4.north|-0,0);%
\ifFnum
  \node[free, anchor=west, text=cgreen] at (13.62,-0.60) {admitted 6.6};%
\else
  \node[free, anchor=west, text=cgreen] at (13.62,-0.60) {admitted};%
\fi
\draw[drop] (g1.south) -- (7.06706,-2.78);%
\draw[drop] (g2.south) -- (9.20219,-2.78);%
\draw[drop] (g3.south) -- (11.33731,-2.78);%
\draw[draw=corange!88!black, line width=0.7pt, densely dashed]
  (7.06706,-2.78) -- (11.33731,-2.78);%
\draw[drop] (9.20219,-2.78) -- (9.20219,-2.9455);%
\node[dat, draw=corange!88!black, fill=corange!7, minimum width=1.60cm,
      minimum height=0.34cm, inner sep=1.2pt] (sink) at (9.20219,-3.20)
  {\Fedg{discarded}};%
\node[free, anchor=west] at (0,-2.42)
  {\ChipL{pyannote}\Cs \Fedg{four increasingly permissive}};%
\node[free, anchor=west] at (0,-2.68)
  {\Fedg{configurations (Table~\ref{tab:vad-levels})}};%
\ifFnum
  \node[free, text=cgray] at (5.9045,-0.45) {725.2};%
  \node[free, text=cgray] at (8.13713,-0.45) {634.5};%
  \node[free, text=cgray] at (10.27225,-0.45) {97.5};%
  \node[free, text=cgray] at (12.40738,-0.45) {24.8};%
\fi
\begin{scope}[shift={(14.71,-2.3655)}]%
  \draw[draw=cgray, line width=0.4pt, rounded corners=1.2pt] (0,0) rectangle (2.99,1.28);%
  \draw[rounded corners=1.2pt, draw=cgreen, line width=0.7pt, fill=cgreen!7]
    (0.12,0.98) rectangle (0.54,1.18);%
  \node[free, anchor=west] at (0.66,1.08) {CAST processing stage};%
  \draw[rounded corners=0.7pt, draw=cnavy, line width=0.4pt, fill=cnavy!8]
    (0.12,0.70) rectangle (0.54,0.90);%
  \node[free, anchor=west] at (0.66,0.80) {Local model};%
  \draw[rounded corners=0.7pt, draw=cred, line width=0.4pt, fill=cred!5, densely dashed]
    (0.12,0.42) rectangle (0.54,0.62);%
  \node[free, anchor=west] at (0.66,0.52) {External API};%
  \draw[draw=corange!88!black, line width=0.7pt, densely dashed] (0.12,0.24) -- (0.54,0.24);%
  \node[free, anchor=west] at (0.66,0.24) {Rejected};%
\end{scope}%
\end{scope}%
\end{tikzpicture}%

%% file: figures/vad_admission_body.tex
\def\Fbarmax{6.05}%
\def\Fbarh{0.30}%
\def\Fpitch{0.46}%
\begin{tikzpicture}[x=1cm, y=1cm, inner sep=0pt, outer sep=0pt]%
\foreach \gv in {300,600,900}{%
  \draw[draw=cink!15, line width=0.3pt]%
    ({\Fbarmax*\gv/900}, 0.22) -- ({\Fbarmax*\gv/900}, {-4*\Fpitch-0.34});%
}%
\foreach \i/\mv/\lbl/\fc/\ec in {%
  0/725.2/{Terminal residuals}/{cgreen!28}/{cgreen!70},%
  1/634.5/{Duration window}/{cgreen}/{cgreen!65!black},%
  2/97.5/{Intelligible word}/{cgreen}/{cgreen!65!black},%
  3/24.8/{Speech check}/{cgreen}/{cgreen!65!black},%
  4/6.6/{Trim to burst}/{cgreen}/{cgreen!65!black}%
}{%
  \fill[fill=\fc, draw=\ec, line width=0.4pt]%
    (0, {-\i*\Fpitch-\Fbarh/2}) rectangle ({\Fbarmax*\mv/900}, {-\i*\Fpitch+\Fbarh/2});%
  \node[anchor=east, text=cink] at (-0.10, {-\i*\Fpitch}) {\Fedg{\lbl}};%
  \node[anchor=west, text=cink] at ({\Fbarmax*\mv/900+0.09}, {-\i*\Fpitch}) {\Fedg{\mv}};%
}%
\draw[draw=cink, line width=0.4pt] (0, {-4*\Fpitch-0.34}) -- (\Fbarmax, {-4*\Fpitch-0.34});%
\foreach \gv in {0,300,600,900}{%
  \draw[draw=cink, line width=0.4pt]%
    ({\Fbarmax*\gv/900}, {-4*\Fpitch-0.34}) -- ({\Fbarmax*\gv/900}, {-4*\Fpitch-0.42});%
  \node[anchor=north, text=cink] at ({\Fbarmax*\gv/900}, {-4*\Fpitch-0.46}) {\Fedg{\gv}};%
}%
\node[anchor=north, text=cink] at ({\Fbarmax/2}, {-4*\Fpitch-0.74})%
  {\Fedg{Residual audio (min)}};%
\end{tikzpicture}%

%% file: figures/language_filter_body.tex
\def\Lbarmax{6.80}%
\def\Lbarh{0.26}%
\def\Lpitch{0.40}%
\def\Laxis{-2.70}%
\begin{tikzpicture}[x=1cm, y=1cm, inner sep=0pt, outer sep=0pt]%
\foreach \gv in {1,2,3,4,5}{%
  \draw[draw=cink!15, line width=0.3pt]%
    ({\Lbarmax*\gv/5}, 0.20) -- ({\Lbarmax*\gv/5}, \Laxis);%
}%
\foreach \i/\mv/\lbl/\fc/\ec in {%
  0/4.55/{Spanish}/{corange}/{corange!65!black},%
  1/0.36/{French}/{corange}/{corange!65!black},%
  2/0.29/{Hawaiian}/{corange}/{corange!65!black},%
  3/0.28/{Portuguese}/{corange}/{corange!65!black},%
  4/0.22/{Manx}/{corange}/{corange!65!black},%
  5/0.22/{Italian}/{corange}/{corange!65!black},%
  6/2.40/{Other labels}/{corange!28}/{corange!70}%
}{%
  \fill[fill=\fc, draw=\ec, line width=0.4pt]%
    (0, {-\i*\Lpitch-\Lbarh/2}) rectangle ({\Lbarmax*\mv/5}, {-\i*\Lpitch+\Lbarh/2});%
  \node[anchor=east, text=cink] at (-0.10, {-\i*\Lpitch}) {\Fedg{\lbl}};%
  \node[anchor=west, text=cink] at ({\Lbarmax*\mv/5+0.09}, {-\i*\Lpitch}) {\Fedg{\mv}};%
}%
\draw[draw=cink, line width=0.4pt] (0, \Laxis) -- (\Lbarmax, \Laxis);%
\foreach \gv in {0,1,2,3,4,5}{%
  \draw[draw=cink, line width=0.4pt]%
    ({\Lbarmax*\gv/5}, \Laxis) -- ({\Lbarmax*\gv/5}, {\Laxis-0.08});%
  \node[anchor=north, text=cink] at ({\Lbarmax*\gv/5}, {\Laxis-0.12}) {\Fedg{\gv}};%
}%
\node[anchor=north, text=cink] at ({\Lbarmax/2}, {\Laxis-0.40})%
  {\Fedg{Removed audio (min)}};%
\end{tikzpicture}%

%% file: figures/lyric_routing_body.tex
\def\Mplotw{7.46}%
\def\Mploth{1.95}%
\def\Mbarw{0.72}%
\def\Mymax{60}%
\begin{tikzpicture}[x=1cm, y=1cm, inner sep=0pt, outer sep=0pt]%
\foreach \gv in {20,40,60}{%
  \draw[draw=cink!15, line width=0.3pt]%
    (0, {\Mploth*\gv/\Mymax}) -- (\Mplotw, {\Mploth*\gv/\Mymax});%
}%
\foreach \i/\av/\cat/\fc/\ec in {%
  0/55/{0}/{cgreen!28}/{cgreen!70},%
  1/27/{1}/{cgreen}/{cgreen!65!black},%
  2/8/{2}/{cgreen}/{cgreen!65!black},%
  3/4/{3}/{cgreen}/{cgreen!65!black},%
  4/6/{4+}/{cgreen}/{cgreen!65!black}%
}{%
  \fill[fill=\fc, draw=\ec, line width=0.4pt]%
    ({\Mplotw*(\i+0.5)/5-\Mbarw/2}, 0)%
    rectangle ({\Mplotw*(\i+0.5)/5+\Mbarw/2}, {\Mploth*\av/\Mymax});%
  \node[anchor=south, text=cink] at%
    ({\Mplotw*(\i+0.5)/5}, {\Mploth*\av/\Mymax+0.04}) {\Fedg{\av}};%
  \node[anchor=north, text=cink] at ({\Mplotw*(\i+0.5)/5}, -0.10) {\Fedg{\cat}};%
}%
\draw[draw=cink, line width=0.4pt] (0, 0) -- (\Mplotw, 0);%
\draw[draw=cink, line width=0.4pt] (0, 0) -- (0, \Mploth);%
\foreach \gv in {0,20,40,60}{%
  \draw[draw=cink, line width=0.4pt]%
    (0, {\Mploth*\gv/\Mymax}) -- (-0.08, {\Mploth*\gv/\Mymax});%
  \node[anchor=east, text=cink] at (-0.12, {\Mploth*\gv/\Mymax}) {\Fedg{\gv}};%
}%
\node[anchor=south, rotate=90, text=cink] at (-0.52, {\Mploth/2}) {\Fedg{Assets}};%
\node[anchor=north, text=cink] at ({\Mplotw/2}, -0.42)%
  {\Fedg{Lyric-tagged spans per asset}};%
\end{tikzpicture}%

%% file: figures/repair_cascade_body.tex
\begin{tikzpicture}[%
  x=1cm, y=1cm,%
  font=\fontsize{7}{8.4}\selectfont,%
  rstage/.style={rounded corners=1.5pt, draw=cgreen, fill=cgreen!16,%
                 line width=0.6pt, text=cink, align=center,%
                 inner xsep=3pt, inner ysep=3.2pt, minimum width=3.20cm},%
  rdata/.style={rounded corners=1.5pt, draw=cgray, fill=cgray!20,%
                line width=0.5pt, text=cink, align=center,%
                inner xsep=3pt, inner ysep=3.2pt, minimum width=3.20cm},%
  rout/.style={rounded corners=1.5pt, draw=cgray!85, fill=cgray!8,%
               line width=0.5pt, text=cink, align=center,%
               inner xsep=3pt, inner ysep=3.0pt, minimum width=3.05cm},%
  rfall/.style={rout, draw=corange, fill=corange!20},%
  flow/.style={-{Latex[length=1.5mm,width=1.2mm]}, draw=cink,%
               line width=0.5pt},%
  side/.style={-{Latex[length=1.5mm,width=1.2mm]}, draw=cgray,%
               line width=0.5pt},%
  fall/.style={-{Latex[length=1.5mm,width=1.2mm]}, draw=corange,%
               line width=0.5pt, densely dashed},%
  elab/.style={text=cink, inner sep=1.5pt, fill=white}%
]%
\node[rdata]  (in)  at (1.72,  0.00) {\textbf{9{,}001} input segments};%
\node[rstage] (det) at (1.72, -1.05) {\textbf{Repetition detector}};%
\node[rstage] (rty) at (1.72, -2.35) {\textbf{Full-window retry}};%
\node[rstage] (spl) at (1.72, -3.65) {\textbf{Split search}};%
\node[rout]  (nf)  at (6.08, -1.05) {8{,}946 not flagged};%
\node[rout]  (p1)  at (6.08, -2.35) {\textbf{37} pass};%
\node[rout]  (p2)  at (6.08, -3.65) {\textbf{14} pass};%
\node[rfall] (bad) at (6.08, -4.80) {\textbf{4} not repaired};%
\draw[flow] (in.south) -- (det.north);%
\draw[flow] (det.south) -- node[elab, pos=0.52] {\textbf{55} flagged (0.61\%)}%
            (rty.north);%
\draw[flow] (rty.south) -- node[elab, pos=0.52] {\textbf{18} still flagged}%
            (spl.north);%
\draw[side] (det.east) -- (nf.west);%
\draw[side] (rty.east) -- (p1.west);%
\draw[side] (spl.east) -- (p2.west);%
\draw[fall] (spl.east) -- (bad.west);%
\end{tikzpicture}%

%% file: main.bbl
\begin{thebibliography}{23}
\providecommand{\natexlab}[1]{#1}

\bibitem[{Ahn et~al.(2026)Ahn, Chae, Park, and Shim}]{whispercd2026}
Ahn, H.; Chae, J.; Park, Y.; and Shim, K. 2026.
\newblock {Whisper-CD}: Accurate Long-Form Speech Recognition using Multi-Negative Contrastive Decoding.
\newblock In \emph{Proc. Interspeech 2026}.
\newblock ArXiv:2603.06193.

\bibitem[{Bain et~al.(2023)Bain, Huh, Han, and Zisserman}]{bain2023whisperx}
Bain, M.; Huh, J.; Han, T.; and Zisserman, A. 2023.
\newblock {WhisperX}: Time-Accurate Speech Transcription of Long-Form Audio.
\newblock In \emph{Interspeech 2023}, 4489--4493.

\bibitem[{Bell et~al.(2015)Bell, Gales, Hain, Kilgour, Lanchantin, Liu, McParland, Renals, Saz, Wester, and Woodland}]{bell2015mgb}
Bell, P.; Gales, M. J.~F.; Hain, T.; Kilgour, J.; Lanchantin, P.; Liu, X.; McParland, A.; Renals, S.; Saz, {\'O}.; Wester, M.; and Woodland, P.~C. 2015.
\newblock The {MGB} Challenge: Evaluating Multi-Genre Broadcast Media Recognition.
\newblock In \emph{2015 IEEE Workshop on Automatic Speech Recognition and Understanding ({ASRU})}, 687--693. IEEE.

\bibitem[{Cettolo et~al.(2026)Cettolo, Cattoni, Negri, and Bentivogli}]{cettolo2026fbk}
Cettolo, M.; Cattoni, R.; Negri, M.; and Bentivogli, L. 2026.
\newblock The {FBK} Sentence-Aware Subtitling System at the {IWSLT} 2026 Subtitling Track.
\newblock In Salesky, E.; Anastasopoulos, A.; Negri, M.; and Federico, M., eds., \emph{Proceedings of the 23rd International Conference on Spoken Language Translation (IWSLT 2026)}, 68--77.
\newblock ISBN 979-8-89176-411-8.

\bibitem[{Chaudhuri et~al.(2018)Chaudhuri, Roth, Ellis, Gallagher, Kaver, Marvin, Pantofaru, Reale, Reid, Wilson, and Xi}]{chaudhuri2018ava}
Chaudhuri, S.; Roth, J.; Ellis, D. P.~W.; Gallagher, A.; Kaver, L.; Marvin, R.; Pantofaru, C.; Reale, N.; Reid, L.~G.; Wilson, K.; and Xi, Z. 2018.
\newblock {AVA-Speech}: A Densely Labeled Dataset of Speech Activity in Movies.
\newblock In \emph{Proc. Interspeech 2018}.

\bibitem[{Davitti et~al.(2024)Davitti, Sandrelli, Korybski, Zou, Orasan, and Braun}]{jat2024asr}
Davitti, E.; Sandrelli, A.; Korybski, T.; Zou, Y.; Orasan, C.; and Braun, S. 2024.
\newblock Using {ASR} Tools to Produce Automatic Subtitles for {TV} Broadcasting: A Cross-Linguistic Comparative Analysis.
\newblock \emph{Journal of Audiovisual Translation}, 7(2): 1--35.

\bibitem[{D{\'e}fossez et~al.(2019)D{\'e}fossez, Usunier, Bottou, and Bach}]{defossez2019demucs}
D{\'e}fossez, A.; Usunier, N.; Bottou, L.; and Bach, F. 2019.
\newblock Demucs: Deep extractor for music sources with extra unlabeled data remixed.
\newblock \emph{arXiv preprint arXiv:1909.01174}.

\bibitem[{Del~Rio et~al.(2022)Del~Rio, Ha, McNamara, Miller, and Chandra}]{delrio2022earnings22}
Del~Rio, M.; Ha, P.; McNamara, Q.; Miller, C.; and Chandra, S. 2022.
\newblock {Earnings-22}: A Practical Benchmark for Accents in the Wild.
\newblock arXiv:2203.15591.

\bibitem[{Gaido et~al.(2021)Gaido, Negri, Cettolo, and Turchi}]{gaido2021beyond}
Gaido, M.; Negri, M.; Cettolo, M.; and Turchi, M. 2021.
\newblock Beyond Voice Activity Detection: Hybrid Audio Segmentation for Direct Speech Translation.
\newblock In \emph{Proceedings of the 4th International Conference on Natural Language and Speech Processing (ICNLSP 2021)}, 55--62. Trento, Italy: Association for Computational Linguistics.

\bibitem[{Kendall and Farrington(2021)}]{kendall2021coraal}
Kendall, T.; and Farrington, C. 2021.
\newblock The Corpus of Regional African American Language.
\newblock Version 2021.07, The Online Resources for African American Language Project, Eugene, OR.

\bibitem[{Koenecke et~al.(2024)Koenecke, Choi, Mei, Schellmann, and Sloane}]{koenecke2024careless}
Koenecke, A.; Choi, A. S.~G.; Mei, K.~X.; Schellmann, H.; and Sloane, M. 2024.
\newblock Careless {Whisper}: Speech-to-Text Hallucination Harms.
\newblock In \emph{Proceedings of the 2024 ACM Conference on Fairness, Accountability, and Transparency}, 1672--1681. New York, NY, USA: Association for Computing Machinery.

\bibitem[{Lan et~al.(2026)Lan, Wei, Guo, Luo, Shang, Li, Rao, Yang, Wu, Huang, and He}]{lan2026hwtsc}
Lan, X.; Wei, D.; Guo, J.; Luo, Y.; Shang, H.; Li, Z.; Rao, Z.; Yang, J.; Wu, Z.; Huang, B.; and He, Y. 2026.
\newblock {HW}-{TSC}'s Submission to the {IWSLT} 2026 Subtitling Track.
\newblock In \emph{Proceedings of the 23rd International Conference on Spoken Language Translation (IWSLT 2026)}, 91--96. San Diego, USA (in-person and online): Association for Computational Linguistics.

\bibitem[{Lin et~al.(2024)Lin, Chen, Zeng, Chen, and Cai}]{apsipa2024cascade}
Lin, Z.; Chen, Z.; Zeng, B.; Chen, L.; and Cai, J. 2024.
\newblock Performance Optimization in the Cascade of {VAD} and {ASR} Systems: A Study on Evaluation and Alignment Strategies.
\newblock In \emph{2024 Asia Pacific Signal and Information Processing Association Annual Summit and Conference (APSIPA ASC)}, 1--6. IEEE.

\bibitem[{Liu et~al.(2025)Liu, Ehrenberg, Lo, Denoix, Barreau, Lample, Delignon, Chandu, von Platen et~al.}]{liu2025voxtral}
Liu, A.~H.; Ehrenberg, A.; Lo, A.; Denoix, C.; Barreau, C.; Lample, G.; Delignon, J.-M.; Chandu, K.~R.; von Platen, P.; et~al. 2025.
\newblock Voxtral.
\newblock arXiv:2507.13264.

\bibitem[{Lucca, Corso, and Pierri(2025)}]{lucca2025subtitles}
Lucca, A.; Corso, F.; and Pierri, F. 2025.
\newblock From Speech to Subtitles: Evaluating {ASR} Models in Subtitling Italian Television Programs.
\newblock Accepted at ECML PKDD 2026, arXiv:2512.19161.

\bibitem[{Papi et~al.(2023)Papi, Gaido, Karakanta, Cettolo, Negri, and Turchi}]{papi2023direct}
Papi, S.; Gaido, M.; Karakanta, A.; Cettolo, M.; Negri, M.; and Turchi, M. 2023.
\newblock Direct Speech Translation for Automatic Subtitling.
\newblock \emph{Transactions of the Association for Computational Linguistics}, 11: 1355--1376.

\bibitem[{Puvvada et~al.(2024)Puvvada, {\.Z}elasko, Huang, Hrinchuk, Koluguri, Dhawan, Majumdar, Rastorgueva, Chen, Lavrukhin, Balam, and Ginsburg}]{puvvada2024canary}
Puvvada, K.~C.; {\.Z}elasko, P.; Huang, H.; Hrinchuk, O.; Koluguri, N.~R.; Dhawan, K.; Majumdar, S.; Rastorgueva, E.; Chen, Z.; Lavrukhin, V.; Balam, J.; and Ginsburg, B. 2024.
\newblock Less is More: Accurate Speech Recognition and Translation without Web-Scale Data.
\newblock In \emph{Proc. Interspeech 2024}.

\bibitem[{Radford et~al.(2023)Radford, Kim, Xu, Brockman, Mcleavey, and Sutskever}]{radford2023whisper}
Radford, A.; Kim, J.~W.; Xu, T.; Brockman, G.; Mcleavey, C.; and Sutskever, I. 2023.
\newblock Robust Speech Recognition via Large-Scale Weak Supervision.
\newblock In Krause, A.; Brunskill, E.; Cho, K.; Engelhardt, B.; Sabato, S.; and Scarlett, J., eds., \emph{Proceedings of the 40th International Conference on Machine Learning}, volume 202 of \emph{Proceedings of Machine Learning Research}, 28492--28518. PMLR.

\bibitem[{Syed et~al.(2025)Syed, Meresman~Higgs, C{\'i}fka, and Sandler}]{syed2025mss}
Syed, J.; Meresman~Higgs, I.; C{\'i}fka, O.; and Sandler, M. 2025.
\newblock Exploiting Music Source Separation for Automatic Lyrics Transcription with {W}hisper.
\newblock In \emph{2025 IEEE International Conference on Multimedia and Expo Workshops (ICMEW)}. IEEE.

\bibitem[{Tsiamas et~al.(2022)Tsiamas, G{\'a}llego, Fonollosa, and Costa-juss{\`a}}]{tsiamas2022shas}
Tsiamas, I.; G{\'a}llego, G.~I.; Fonollosa, J. A.~R.; and Costa-juss{\`a}, M.~R. 2022.
\newblock {SHAS}: Approaching Optimal Segmentation for End-to-End Speech Translation.
\newblock In \emph{{Interspeech 2022}}, 106--110.

\bibitem[{Wang et~al.(2025)Wang, Alhmoud, Alsahly, Alqurishi, and Ravanelli}]{wang2025calm}
Wang, Y.; Alhmoud, A.; Alsahly, S.; Alqurishi, M.; and Ravanelli, M. 2025.
\newblock {Calm-Whisper}: Reduce {W}hisper Hallucination on Non-Speech by Calming Crazy Heads Down.
\newblock In \emph{{Interspeech 2025}}, 3414--3418.

\bibitem[{Wilken, Georgakopoulou, and Matusov(2022)}]{wilken2022suber}
Wilken, P.; Georgakopoulou, P.; and Matusov, E. 2022.
\newblock {S}ub{ER} -- A Metric for Automatic Evaluation of Subtitle Quality.
\newblock In \emph{Proceedings of the 19th International Conference on Spoken Language Translation (IWSLT 2022)}, 1--10. Dublin, Ireland (in-person and online): Association for Computational Linguistics.

\bibitem[{Xu et~al.(2023)Xu, Jia, Majumdar, Huang, Watanabe, and Ginsburg}]{xu2023tdt}
Xu, H.; Jia, F.; Majumdar, S.; Huang, H.; Watanabe, S.; and Ginsburg, B. 2023.
\newblock Efficient Sequence Transduction by Jointly Predicting Tokens and Durations.
\newblock In \emph{Proceedings of the 40th International Conference on Machine Learning}, volume 202 of \emph{Proceedings of Machine Learning Research}, 38462--38484. PMLR.

\end{thebibliography}
